\documentclass{article}

\usepackage{epsfig}
\usepackage{subcaption}
\usepackage{calc}
\usepackage{amssymb}
\usepackage{amstext}
\usepackage{amsmath}
\usepackage{amsthm}
\usepackage{multicol}
\usepackage{pslatex}
\usepackage{apalike}
\usepackage{tikz}
\usepackage{array}
\usepackage{authblk}

\providecommand{\keywords}[1]{\textbf{\textit{Keywords---}} #1}

\begin{document}

\title{Temporal Motifs in Smart Grid}

\author{
  Rucha Bhalchandra Joshi \\
  \texttt{rucha.joshi@niser.ac.in}
  \and
  Annada Prasad Behera\\
  \texttt{annada.behera@niser.ac.in}
  \and
  Subhankar Mishra\\
  \texttt{smishra@niser.ac.in}
}

\affil{Machine Learning And Building (MLAB), \\School of Computer Sciences \\ National Institute of Science Education and Research, Bhubaneswar - 752050, India\\%
    Homi Bhabha National Institute, Anushaktinagar, Mumbai - 400094, India}
    
\date{}

\maketitle

\abstract{
A complex network can be characterized by patterns. Such frequently occurring significant patterns are called motifs and in a time dependent network, they are called temporal motifs. One of the temporal networks where temporal motifs are observed and play a major role; is the Smart Grid. The energy consumption pattern across the appliances, houses, communities and entire cities help energy utility companies and consumers plan their electricity generation and consumption. The temporal motifs for the smart grid constitutes of the consumers and producers and the edge or connection represents energy flow between two participants of the network, these connections last till the power is being consumed/generated. This paper formally defines the temporal motifs for smart grid network and proposes a way to create such temporal motifs in the network. We also discuss how the temporal motifs fit into the hierarchical structure of power distribution system of Smart Grid. }

\keywords{Smart Grid, Temporal Motifs, Complex Systems, Cyber-physical Systems}



\section{\uppercase{Introduction}}
\label{sec:introduction}

\noindent Many complex systems can be abstracted with the help of networks. 
The entities participating in the systems are modeled as nodes and the 
relations by which they are linked to each other are modeled as the 
edges of a network graph. Abstractions help us study the complex system such as food chain, citation network. Time dependent systems can be abstracted as temporal networks. Some notable examples of temporal network are Facebook, Email as well as recent networks such as Bitcoin. Structure of the temporal network changes with time. Since the 
edges in temporal graphs depend on time, their presence is determined only 
at a given time. To understand the behaviour of the temporal network, it 
is essential to consider the time of occurrence of temporal edges. 

One such system is smart grid network. A smart grid has various entities, such as producers, consumers, transmitters of power, participating in the network. The hierarchical structure of the smart grid has been discussed by \cite{aggarwal_proposed_2010,mishra2016price} where the power distribution in the grid is according to the voltage. \cite{rech_towards_2012} discussed the existence of the consumption sector of the hierarchy wherein the lowermost layer consists of the last links of distribution grid connecting consumers to the main grid. A transformer supplies the power to the consumers that are connected to it. So, the second layer from the bottom consists of transformers. The third layer is of substations that supply power to the transformers in second layer. This hierarchy goes up along with all the participating entities in the grid network. Pattern of distribution of power consumption over the period of time form motifs. These motifs depicts the behaviour of the participating entities.

In a smart grid, the meters can get the data pertaining 
to each of the appliances used in the house that consumes amount of electricity. Smart meter keeps track of the consumption of each of the rooms or any other infrastructure 
that consumes electricity. Hence, at household level, the electricity 
consumption and distribution is known with the help of smart meters. The inferences about the usage of electricity at the lowermost layer in a smart grid can be drawn by taking this consumption data into consideration.

Our contributions are as follows:
\begin{itemize}
    \item We formally define temporal motifs that occur in smart grid network.
    \item We show a method for constructing such temporal motifs in smart grid network.
    \item We discuss temporal motifs in smart grid with overlapping window and fitting the temporal motifs in the hierarchical structure of the smart grid.
\end{itemize}

The remainder of the paper is organized as follows: Section 2 discusses the related work. We discuss the background work in Section 3. Section 4 gives the detailed explanation of our proposed model. We present a case study and discussion of temporal motifs in smart grid network are in Section 3 and 4 respectively.  We present the conclusions in Section 6.

\section{\uppercase{Related Work}}

\noindent Motifs \cite{milo_network_2002} are the basic building blocks of a complex network.
They are recurring, significant patterns of interconnections. The network motifs are defined to study the structural design principles of complex networks in various fields such as biochemistry, neurobiology, ecology, engineering and so on.

Motifs, defined as the frequently occurring and significant patterns in time, can be used to characterize the time series data \cite{Lin_SAX_2002}. Motifs in temporal networks have been defined in order to understand their role in the temporal networks such as the network of emails, phone calls, social media etc. \cite{Paranjape_2017}. However, they only consider the occurrence of the edges one at a time without any duration attached to the existence of the edges.

Motif based pattern detection technique was proposed to discover regular behaviour of smart meter users \cite{funde_motif-based_2018}. The model proposed by \cite{funde_motif-based_2018} considers one appliance at a time and detects the motifs formed by it. They develop temporal association rule mining to find the relation between usage of energy by various appliances in a particular time period. However, considering only one appliance at a time does not tell a complete story about the consumption pattern of the members of the house.

\section{\uppercase{Background Work}}

\subsection{Temporal Graph and Motif}
In a temporal graph $G = (V, E)$ where $V$ is set of vertices and $E$ is set of edges, the temporal edges are represented as $(u, v, t)$ where $u, v \in V$ and a timestamp $t$ is associated with the edge. Temporal motif is a collection of edges in a particular sequence that form a particular structure in a given time window $\delta$. Since the timestamps are attached with each of the edges in temporal network, the motif is the structure occurring within time $\delta$ from the occurrence time of first edge. This time window of size $\delta$ is slid over the time as we consider the next motif.
When we consider the temporal edges in smart grid, the edges occur at different times but they last for a duration of time. Smart grid is a specific application of temporal network where the edges have a time of occurrence and it remains in the network till the appliance that caused it to occur was switched off or no longer draws any energy from the meter. This differentiates our work from the previous work as edges stay alive for a certain duration. We say that an edge occurs at a given time when the appliance corresponding to the node connecting the edge is turned on. The smart meter captures the energy at particular time interval in a sequence, hence we have to assign a window time to it rather than the actual start time of the consumption. The energy consumption by same appliance may vary at different times. 


\subsection{Topology of Smart Grid}
\begin{figure}[h]
    \centering
    \includegraphics[width=7.5cm,height=10cm,keepaspectratio]{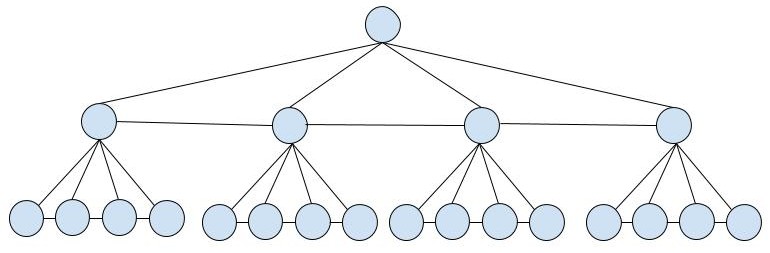}
    \caption{Topology of Power Grid}
    \label{fig:topology}
\end{figure}

The power distribution grid is arranged according to the voltage \cite{aggarwal_proposed_2010}. The various levels of the hierarchy are connected using the voltage networks, here power plants are connected via high voltage networks and the level of household appliances is connected using a voltage network. The smart grid is defined to consist of nodes $N$ and interconnected edges $E$; where nodes represent the actors and are connected to $\#b$ other nodes ($b$ is branching factor). The levels of this hierarchy is denoted by $L$ \cite{rech_towards_2012}.

\section{\uppercase{Smart Grid Temporal Motifs}}
\begin{figure}[h]
    \centering
    \includegraphics[width=8cm,height=10cm,keepaspectratio]{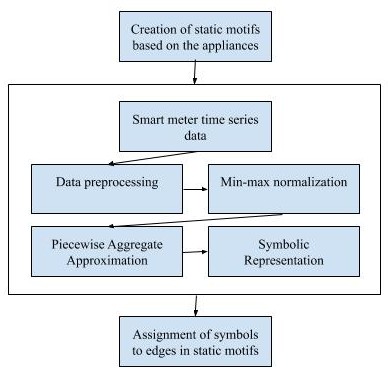}
    \caption{Overview of motif creation process for smart grids}
    \label{fig:model}
\end{figure}

\noindent We propose a model (Figure \ref{fig:model}) for creation of temporal star motifs with associated symbol corresponding to the energy consumption. We also show how these motifs help to draw inferences from the hierarchy of the participants of a smart grid network. In our proposed model the edges of the motifs have been tagged with symbols associated with corresponding energy consumption levels and time window. These timestamps are indicatives of window frame numbers. They imply the order of occurrences of the motif structures.

Consider $m_i$ to be meter reading (total) at a given time $i$. We consider a time series $T = {m_1, m_2, ... , m_t}$. Each $m_i$ consists of the values corresponding to internal distribution of energy among all the appliances utilizing the energy at a given time $i$. Let $A$ be the set of appliances. Let $c^j_i$ be the energy consumed by the $j^{th}$ appliance at time $i$. Therefore,


 $$m_i = \sum_{j \in A} c^j_i $$


\subsection{Motif Creation}

\begin{figure}[ht]
    \centering
    \includegraphics[width=5cm,height=10cm,keepaspectratio]{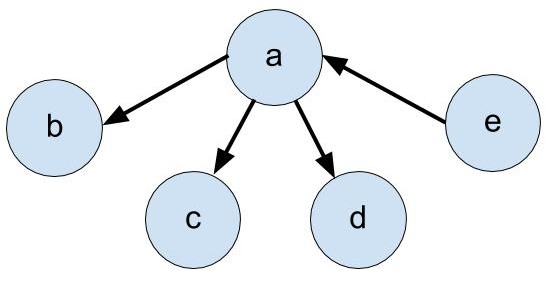}
    \caption{Example of a star motif. There are 5 nodes in total. One center node and four neighbors to the center node. The direction of the edges between the center node and the neighboring nodes depends on the relationship between them.}
    \label{fig:star-motif}
\end{figure}

A star motif is defined as a graph of $k$ nodes in total, out which one node (the center node) has $k-1$ neighbors, which is the center node while all the other nodes have only one neighbor each. An example of star motif is shown in Fig. \ref{fig:star-motif}.

\begin{figure}[h]
    \centering
    \includegraphics[width=7cm,height=10cm,keepaspectratio]{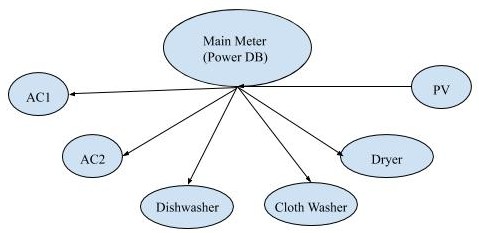}
    \caption{Example of a static star motif in a house}
    \label{fig:star-motif-house}
\end{figure}
We consider a particular consumer where there are various appliances in a house that contribute to overall consumption of energy in the house. All of which are essentially connected to the main smart meter. We create a star motif of these appliances along with the smart meter. Fig \ref{fig:star-motif-house} shows a star motif within a house where we consider the smart meter in the house to be the center node and all the appliances to be the rest of the nodes which are only connected to the center node. The nodes for the appliances that draw energy from the meter are represented using an edge from meter (center node) to the appliance (corresponding neighboring node). For any other equipment that generates energy (e.g. solar panel), the edge goes from the equipment to the center node. 


\subsection{Symbolic Representation}
\begin{enumerate}
\item\textbf{Data Preprocessing and Min-max Normalization}.
\label{sec:prep}
Time series data preprocessing is done to normalize the data. We consider smart meter data which keeps track of consumption of each of the appliance. We perform min-max normalization so that the values after normalization lies between $[0,1]$. The normalized value $y$, of a data point $x$, is given by
\begin{equation*}
    y = \frac{(x - min)}{(max - min)}
\end{equation*}
where $min$ is the minimum and $max$ is the maximum data value in the dataset.

\item\textbf{Piecewise Aggregate Approximation}.
\noindent On the normalized data, we apply Piecewise Aggregate Approximation (PAA) to discretize the data. By selecting the right parameters in PAA, it can be altered to suit the needs of the application at hand. The normalized time series data is divided into $w$ windows. 
The average of values in every window is calculated.

\item\textbf{Matching Symbols to Energy Levels}.
\noindent After PAA, we represent the energy consumption of each of the appliance with a symbol. Number of energy levels and their corresponding symbols is another parameter that can be set according to use case. The symbol values represent the levels of energy consumption over normalized data. The data values range between 0 and 1, so we decide on number of symbol to be used and range of each of energy consumption corresponding to each symbol. The number of energy levels vary application to application. 

\end{enumerate}

\subsection{Symbol assignments to edges of a motif}
\noindent The edges in static motif that we created in first step, for a house is assigned timestamps associated with the time window in which we are determining the associated symbol. In the static motifs, edges are between the main line and an appliance. The temporal edge occurs in smart grid in the window in which an appliance is turned on.

We define the temporal edge for grid network as quadruple represented as $(u, v, t_w, x)$. Table \ref{tab:descr} describes each component of the quadruple. We perform the operations described in the previous steps on the data, to get the symbols associated for each of the appliance in the same set of time windows. Then, for a particular window, consider all the appliances along with meter as nodes, while the energy production and consumption among appliances determine the directions between the edges and the level of energy consumption is given by the symbol associated with the edge. The time window is in which we are determining the consumption level is given by the timestamp corresponding to the time window.


\begin{table}
    \centering
    \begin{tabular}{ | m{1cm} | m{13em} | } 
            \hline
            Variable & Description \\
            \hline
            \hline
            $u$ & Supplier of power (any appliance that produces or supplies power)\\
            \hline
            $v$ & Consumer of power (any appliance that consumes power) \\ 
             \hline
            $t_w$ & Timestamp of the time window corresponding to the edge \\ 
            \hline
            $x$ & Symbol corresponding to the level of energy consumption assigned to the edge n the given window\\
            \hline
    \end{tabular}
    \caption{Description of variables associated with a temporal edge in smart grid}
    \label{tab:descr}
\end{table}

The complete data of consumption in a grid can be represented as the collection of the temporal edges we defined earlier.

\begin{figure}
    \centering
    \includegraphics[width=7cm,height=10cm,keepaspectratio]{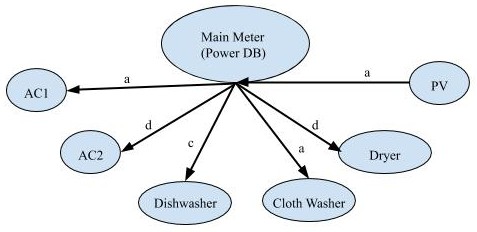}
    \caption{Example of star motif in a house}
    \label{fig:motif-with-symbols}
\end{figure}

\subsection{Temporal Motifs for Electric Grid Network}
As shown in Fig. \ref{fig:motif-with-symbols}, along with assignments of symbols to the edges in static motif, note that this motif structure occurs in a particular time duration, since each of the symbols assigned to the edges represent the level of consumption in a particular time period. The motif helps us to look at the consumption and distribution of energy among all appliances in a time slot in a house. The collection of such motifs over a time windows of size $\delta$ is defined to be a temporal motif for energy consumption data.
\section{\uppercase{Case Study}}
We take an example to demonstrate the steps to create temporal motifs in smart grid network. We take a part of Pecan Street Dataport \cite{data} from 15 minute dataset. We consider a house with data-id 27, which is located at New York. For this house, we consider the consumption values for the time period 04:00:00 to 07.00.00 on 2019-05-01.

\subsection{Motif Creation}
\label{sec:static-motif-create}
\begin{figure} [h]
    \centering
    \includegraphics[width=7cm,height=10cm,keepaspectratio]{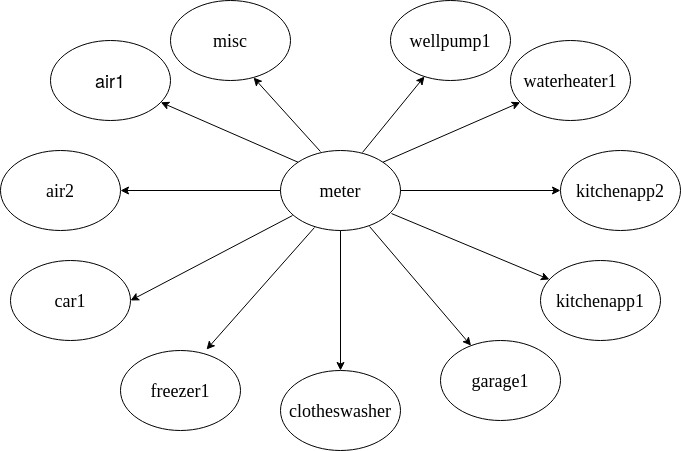}
    \caption{Static Motif for house ID 27}
    \label{fig:motif1}
\end{figure}
The underlying star motif for this house ID is shown in Fig.\ref{fig:motif1}. This is derived based on the appliances used in the house.

\subsection{Symbolic Representation}
\begin{enumerate}
\item\textbf{Data Preprocesssing and Min-max Normalization}.
Preprocessing and min-max normalization of data is done according to formulas mentioned in Section \ref{sec:prep} to get the normalized data. 

\item\textbf{Piecewise Aggregate Approximation}.
Since the data has a duration of 3 hours, with windows size of 1 hour, the number of windows $w = 3$.
For any given time window there is a value attached to it which is average of the values corresponding to the timestamps in the window.

\item\textbf{Matching Symbols to Energy Levels}.
We consider 4 symbols $a, b, c $ and $d$, they correspond to four consumption levels. Very low consumption is represented by a symbol $a$, $b$ represents average energy consumption, $c$ represents more than average consumption of energy while very high energy consumption is represented by symbol $d$. 

We define the range for each of the symbols as shown in Table.\ref{tab:range}.

\begin{table}[h]
    \centering
    \begin{tabular}{ | c | c | } 
            \hline
            Symbol & Range \\
            \hline
            \hline
            $a$ & $0 \leq value < 0.25$ \\
            \hline
            $b$ & $0.25 \leq value < 0.5$\\
            \hline
            $c$ & $0.5 \leq value < 0.75$\\
            \hline
            $d$ & $0.75 \leq value \leq 1$\\
            \hline
    \end{tabular}
    \caption{Range for each of the symbols}
    \label{tab:range}
\end{table}

\end{enumerate}
\subsection{Symbol assignments to edges of a motif}
\label{sec:symbol_assign}
\begin{figure}
    \centering
    \begin{subfigure}{.5\textwidth}
      \centering
      \includegraphics[width=.8\linewidth]{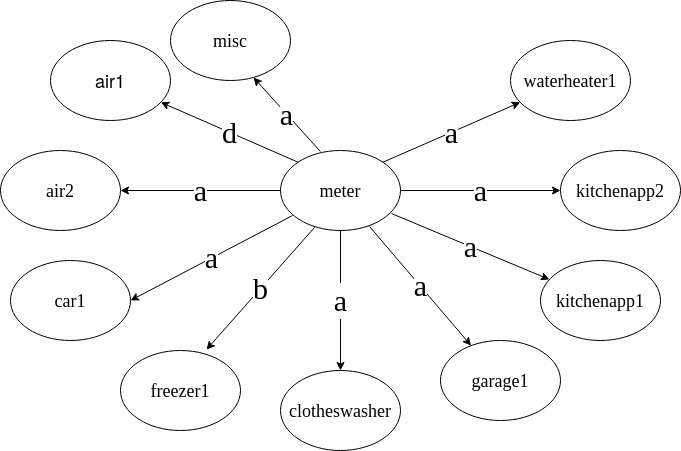}  
      \caption{Motif at time $t_1$}
      \label{fig:motif2}
    \end{subfigure}\vspace{.3in}
    
    \begin{subfigure}{.5\textwidth}
      \centering
      \includegraphics[width=.8\linewidth]{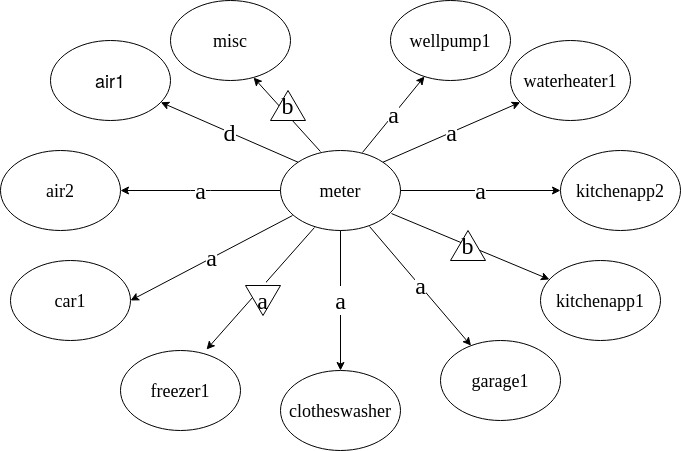}  
      \caption{Motif at time $t_2$}
      \label{fig:motif3}
    \end{subfigure}\vspace{.3in}
    
    \begin{subfigure}{.5\textwidth}
      \centering
      \includegraphics[width=.8\linewidth]{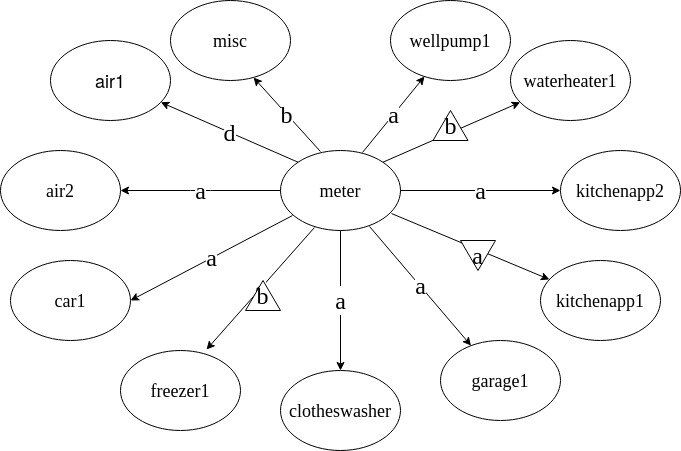}  
      \caption{Motif at time $t_3$}
      \label{fig:motif4}
    \end{subfigure}
    
    \caption{Temporal motifs in smart grid network}
    \label{fig:temp-motif}
\end{figure}

Each edge in the static motif created in Section \ref{sec:static-motif-create} has a symbol that corresponds to the average of the values of consumption on each edge in a window. The symbols are assigned to the edges. An example of a motif in House ID 27 for time window labelled $t_1$ for duration on 1 hour is shown in Fig.\ref{fig:motif2}.

\subsection{Temporal Motifs for Electric Grid Network}.
A sequence of motifs created in Section \ref{sec:symbol_assign} may have different symbols associated with their edges in different time windows, since the energy consumption varies over time. Such a sequence is the temporal motif in a smart grid network. The final temporal motif with 3 time windows is shown in Fig.\ref{fig:temp-motif}. Increase in consumption level is indicated by symbols inside upward pointing triangles, whereas the decrease in consumption is indicated by symbols inside triangles facing downward in the temporal motifs.

\section{\uppercase{Discussion}}
\subsubsection*{Overlapping Temporal Motifs}
\noindent While we only consider the window for finding the symbols associated with edges to be non-overlapping, other possibilities to be considered are overlapping window and the absolute consumption values. Considering absolute values is too specific because the consumption of energy on one day may not be exactly the same as that on the next day. Hence assigning symbols based on absolute values at a given time may not help infer anything useful about the data. In a time period, if the average data consumption is considered then it gives the approximate consumption level of the appliance. 

Overlapping window can be considered if it is needed for the application under consideration. We slide the window over the time duration with some predetermined time overlapping in two consecutive windows. This would help in maintaining the information related to continuity of the data to some extent depending on how much the overlap is.
\subsubsection*{Fitting Temporal Motifs into Smart Grid hierarchy}
We propose this model for residential type of locality. This is easily scalable to other types of localities as well, such as industrial, commercial etc and can be extended to fit other hierarchical levels in the smart grid as well as other complex networks. To study the role that these motifs play in smart grid network, it is essential that we consider the hierarchy of the participants of the network. 
The hierarchy described in \cite{rech_towards_2012} is discussed below to suit for the requirement of our model.
\begin{itemize}
    \item The very basic level in the hierarchy consists of the appliances in a household. As discussed earlier the motifs which are formed at this layer are based on electricity consumption of each of the appliances at various times. 
    \item The layer above the layer of appliances is of houses in a locality or a community residing at a particular location. The motifs would consist of the houses in the locality and the point of supply of electricity to all these houses as nodes. The consumption of each of the houses at various times would determine the edges and the direction of the edges. These motifs can be defined in similar fashion as we did earlier by determining the suitable parameter values. 
    \item Similarly, the next layer consists of communities which together form a city.
    \item More layers can be thought of and considered on top of the previously mentioned layers so as to build the model of motifs that will help enable us to study and determine various aspects of a smart grid network.
\end{itemize}





\section{\uppercase{Conclusion}}
\label{sec:conclusion}

\noindent Temporal motifs play an important role in characterizing networks. The change in usage of power over time helps to study the behaviour of the consumers. We have formally defined the temporal motifs in smart grid network. We have also described the construction of such temporal motifs in smart grid network, without overlapping windows, how to fit them in the hierarchical structure of the network. In future we will use motifs to draw inferences about participants of the grid network. We will also look at impact of this model on the energy distribution policies.


\section*{\uppercase{Acknowledgements}}
This research was partially supported by NRDMS/UG/S.Mishra/Odisha/E-01/2018 

\bibliographystyle{apalike}
{\small
\bibliography{main}}



\end{document}